# Ferromagnetic Mixed-Valence and Kondo Lattice State in TmTe at High Pressure

P. Link,[1] I.N. Goncharenko,[1,2] J.M. Mignot,[1] T. Matsumura,[3] and T. Suzuki[3]

[1] *Laboratoire Léon Brillouin, CEA-CNRS. CEA-Saclay, 91191 Gif sur Yvette, France*
[2] *Russian Research Center 'Kurchatov Institut', 123182 Moscow, Russia*
[3] *Department of Physics, Tohoku University, Sendai 980-77, Japan*

Neutron diffraction experiments on TmTe at pressures up to 7 GPa are reported. The semiconductor-to-metal transition occurring at 2 GPa in this compound is accompanied by the appearance of strong ferromagnetic exchange interactions producing a Curie temperature as high as 14 K. This behavior is characteristic for the incipient mixed-valence regime just above the transition pressure and traced back to the release of a small concentration of free charge carriers in the material. The steep decrease of both the Curie temperature and ordered magnetic moment occurring at higher pressures emphasizes the role of Kondo fluctuations, and raises the interesting possibility that a quantum critical transition to a non-magnetic ground state of mixed-valence Tm might take place around 6 GPa.

In magnetic materials, the concentration of mobile charge carriers is known to strongly influence the nature and strength of exchange interactions between neighboring atoms. This is evidenced in a particularly striking manner by systems undergoing metal-insulator transitions upon doping such as Eu chalcogenides [1] and Eu hexaboride [2], or by La manganites where the existence of hole-mediated "double-exchange" interactions is assumed to be responsible for the orders-of-magnitude response of the electrical resistivity to an applied magnetic field [3]. In the above examples, the extra carriers are usually created by substituting the rare-earth element by an ion with a different valence. This clearly produces a local charge singularity, which can have non-trivial consequences such as e.g. the formation of impurity states, bound magnetic polarons, etc. [4]. A more uniform situation may be achieved in the case where the charge carriers result from a homogeneous mixed-valence state of all magnetic atoms. This phenomenon can be viewed as the quantum admixture of two electronic configurations corresponding to different occupations $n$ and $n-1$ of the rare-earth $4f$-shell, one electron being able to fluctuate between a localized $4f$-orbital and extended $5d$-$6s$ states. Numerous examples of this behavior have been found experimentally, mostly in Ce, Sm, Eu, Tm and Yb compounds [5]. The mixed valence can occur spontaneously, or be the result of an external perturbation, typically chemical substitution or applied pressure. However, in the overwhelming majority of these systems, the valence instability is accompanied by large spin fluctuations which cause a complete suppression of the magnetic state and obscure the variation of magnetic exchange interactions. The only notable exception to date has been reported for the cubic (NaCl structure) thulium monochalcogenide family Tm$X$ ($X$ = S, Se, Te) [6]. Among these materials, stoichiometric TmSe exhibits antiferromagnetic (AF) order below $T_N$ = 3.4 K with a valence estimated to be comprised, depending on the technique used, between 2.55 (lattice constant) and 2.72 (effective moment). This unique behavior has been claimed to result from the fact that both valence states Tm$^{2+}$ and Tm$^{3+}$ have degenerate low-lying multiplets (respectively $^3H_6$ and $^2F_{7/2}$) but, for lack of a reliable solution to the many-body ground state problem, this conclusion should be taken with caution. On the other hand TmTe is a divalent magnetic semiconductor with a gap of about 0.35 eV and very weak exchange interactions. However, an external pressure of 2 GPa is sufficient to close the energy gap and drive the system to a mixed-valence state, with a much higher (albeit not fully metallic) electrical conductivity [7,8]. Whereas Tm$^{2+}$Te at $P$ = 0 orders magnetically (AFII-type) only below 0.5 K [9], clear anomalies detected in the $\rho(T)$ curves under pressure [8] suggest a much higher ordering point (~ 15 K) in the mixed-valence regime. This system therefore represents an attractive case for studying the interplay between electronic transport and magnetic phenomena close to a semiconductor-to-metal transition.

In this Letter, we report the experimental observation, by neutron diffraction up to $P$ = 7 GPa, of a new ferromagnetic (FM) phase associated with the mixed-valence state of TmTe. The pressure dependencies of the Curie temperature and ordered moment are compared with earlier data for Tm(Se,Te) solid solutions [10,11] and similarities with other "low-carrier" systems are emphasized. It is further argued, from the analysis of the temperature and pressure dependence of the order parameter, that strong quantum fluctuations exist in this system, which might lead to a complete suppression of the Tm magnetic moment in a critical pressure region situated around 6 GPa.

The experiments have been carried out on the two-axes lifting detector diffractometer 6T2 at the Orphée neutron facility in Saclay using a neutron wavelength of 0.234 nm. Single-crystals with volumes ranging from 0.01 mm$^3$ for the lowest pressure to 0.003 mm$^3$ for 5.4 GPa were extracted from the same batch already used for high-pressure resistivity measurements [8]. Pressure was generated in a sapphire-anvil cell described elsewhere [12]. The pressure transmitting medium was a 4:1 ethanol-methanol mixture providing almost hydrostatic conditions in this pressure range. The pressure was measured by the ruby-fluorescence method, with an total error smaller than ± 0.2 GPa. The out-of-plane angular range of the detector and the cell geometry make it possible to observe nuclear reflections ($hkl$) with $l=0,1$ and $h,k \leq 3$ for typical lattice parameters of about 0.6 nm. Neutron-diffraction experiments were carried out at 5 different pressures: 2.3, 2.7, 3.7, 4.5 and 5.4 GPa in a standard helium-flow cryostat with a minimum temperature $T_{min} = 1.5$ K.

At $P = 2.3$ GPa, elastic $Q$-scans performed at $T_{min}$ along the main symmetry directions of the reciprocal lattice did not reveal any extra magnetic peaks. In particular, no magnetic intensity could be detected at positions corresponding to $k = (1,0,0)$ (AFI) and $k = (1/2,1/2,1/2)$ (AFII), nor superimposed on the weak nuclear peaks ($h,k,l$ = odd integers). It was thus concluded that at this pressure, the system does not order magnetically down to 1.5 K.

In contrast to this, the data for $P = 2.7$ GPa show clear evidence for FM order, with a Curie temperature $T_C$ as high as 14K. This is demonstrated in the inset of Fig. 1 where rocking curves through the (111) reflection are plotted for $T = 2$ K and $T = 29$ K > $T_C$. Note that the conditions for observing the FM component are particularly favorable for reflections with odd indices due to the near cancellation of the nuclear scattering from Tm and Te. To fully solve the magnetic structure, one needs to determine i) the magnitude of the ordered moment, ii) its direction, and iii) the population of the different magnetic domains. A good fit of the experimental intensities was obtained by assuming the moments to be oriented along the cubic axes, as they are in the AF I phase of TmSe. The weak magnetic contribution to the (200)-type reflections implies that the $z$-domain (moments aligned along the cell axis) is strongly depopulated, which can be ascribed to the weak uniaxial stress components existing within the pressure cell. For this magnetic structure, the intensity of the (111)-type peaks does not depend on domain populations and directly provides the value of the magnetic moment $\mu = 2.1(1) \mu_B$/Tm at $T = 2$ K [13].

For the higher pressures of 3.7, 4.5 and 5.4 GPa, the same FM Bragg peaks as at 2.7 GPa are observed. The temperature dependencies of the intensities for the (111) reflection are summarized in Fig. 1. The Curie temperatures (indicated by arrows) decrease rapidly as a function of pressure. Quite remarkably, the data for each pressure display a linear temperature dependence of the magnetic intensity below the Curie point and down to $T_{min}$ (except in the narrow saturation region reached below ≈ 6 K for $P = 2.7$ and 3.7 GPa). Since the magnetic intensity is proportional to the square of the order parameter, this behavior implies a critical exponent equal to 0.5 in this temperature window, which is characteristic for a mean-field behavior. As the values plotted in the figure have been normalized to the nuclear intensity of the (200) reflection at the same pressure, a strong reduction of the Tm magnetic moment with increasing pressure can be derived directly from the decrease of the magnetic intensity at $T_{min}$ (even though complete saturation is not reached at this temperature for the higher pressures, the uncertainty on the extrapolation to $T = 0$ is not significant). The pressure dependencies of both $T_C$ and $\mu_0$ (defined here as the moment value at $T_{min}$) are plotted in Fig. 2, together with the result of a powder neutron diffraction measurement performed at 7.0 GPa on the multidetector instrument G6-1 at $\lambda = 0.476$ nm. As shown in the inset, the decrease of the Curie temperature is proportional to the square of the magnetic moment between 2.7 and 5.4 GPa, where they reach respectively 4.5 K and 1.2 $\mu_B$. A universal curve can actually be produced by plotting $I(T)/\mu_0^2(P)$ as a function of $T/T_C(P)$. At $P = 7.0$ GPa, the absence of any detectable magnetic intensity in the powder diffraction pattern measured at 1.5 K puts an upper limit of about 1 $\mu_B$ to the value magnetic moment at this temperature. This clearly indicates that the reduction of $T_C$ and/or $\mu_0$ continues beyond 5.4 GPa.

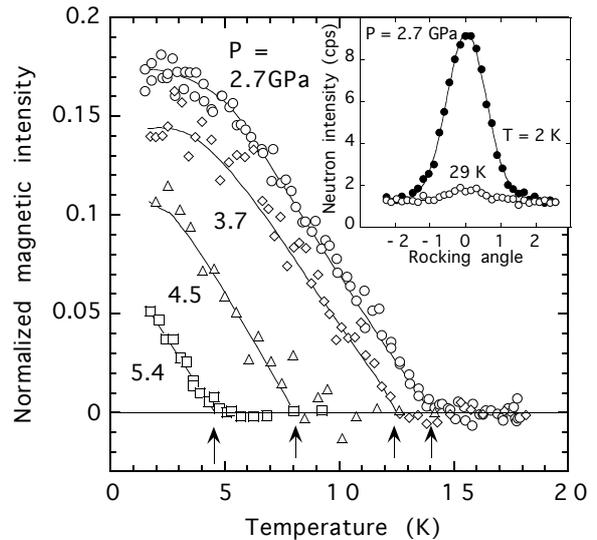

Fig. 1. Temperature dependence of the magnetic contribution to the (111) peak intensity, normalized to the (200) nuclear intensity for different pressures. Arrows mark the estimated Curie-temperatures, lines are guides to the eye. Inset: rocking curves of the (111) reflection at 2 K and 29 K for $P = 2.7$ GPa.

The abrupt jump of the ordering temperature of TmTe from $T_N < 1$ K at 2.3 GPa to $T_C = 14$ K at 2.7 GPa points to a dramatic increase of the FM exchange interactions. A similar behavior was previously reported [10,11] for the solid solution TmSe$_{0.6}$Te$_{0.4}$, but the highest ordering temperature found in that work was only 5.5 K. One explanation for this discrepancy could be the presence of local disorder (e.g. partly inhomogeneous valence mixing) in the alloy systems, due to preferred di- (tri-) valent occupancy of Tm sites having a given number of Te (Se) neighbors. Another difference comes from the fact that, in the compound with only 0.4 Te content, the transformation into the metallic state under pressure takes place through a first-order phase transition, which might make the valence range corresponding to the highest $T_C$ values experimentally inaccessible.

Interestingly, recent high-pressure experiments on EuTe [14] have shown that the ground state of this system changes from AF at ambient pressure to FM above 10 GPa without variation of the valence state. It was suggested that indirect-exchange FM interactions between nearest-neighbor Eu ions, governed by the inter-site hybridization of their $f$- and $d$-orbitals [4,15], are enhanced as interatomic distances are reduced. Such an effect should also exist in TmTe, and indeed it would be interesting to investigate the pressure dependence of its magnetic order between 0 and 2.3 GPa at temperatures below 1 K. However, it should be emphasized that the ferromagnetic onset observed in TmTe practically coincides with the closing of the semiconducting gap (resistivity measurements in Ref. [8]) and the entry into the mixed-valence state (compressibility measurements in Refs. [16] and [17]). Furthermore, $T_C$ is found to decrease steadily from its maximum value, which is reached immediately after the semiconductor-to-metal transition. This behavior is at variance with that of EuTe, and strongly suggests that the FM state realized in TmTe under pressure has a different origin which directly reflects the release of free charge carriers into band electron states.

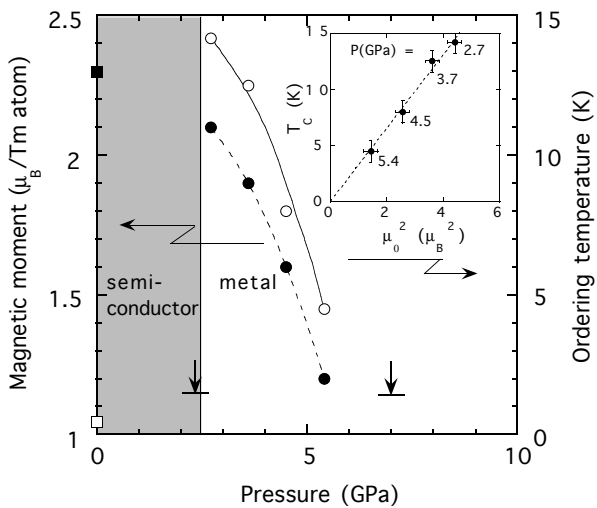

Fig. 2. Pressure dependence of the ordered magnetic moment (●, left scale) and the ordering temperature (○, right scale). The ambient-pressure values are taken from ref. 9. At $P =$ 2.3 GPa and 7 GPa, where no magnetic order has been detected within experimental accuracy ($\mu \leq 1\mu_B$), the vertical arrows indicate the minimum temperature of measurement. The lines through data points are guides to the eye. Inset: proportionality of $T_C$ and $\mu_0^2$.

Back in the 1980's, it was noted in different experimental studies that, in the Tm$X$ ($X$ = S, Se, Te) family of compounds and alloys, the FM character generally increases when the average thulium valence becomes closer to 2+ [6]. Even in the strongly mixed-valent state of pure TmSe, a FM tendency is clearly evidenced by the metamagnetic transition occurring in a weak applied magnetic field $H_M = 0.4$ T [18].

It was then argued by Varma [19] that this behavior results from a double-exchange interaction, similar to that existing in doped La manganites [3]. In this picture, it is assumed that the substitution of Tm$^{2+}$ sites into an originally trivalent Tm$^{3+}X$ matrix results in the formation of sites with one excess 4$f$-electron. Due to strong Hund's rule couplings, the propagation of this electron from site to site favors a FM alignment of the Tm magnetic moments. The free electrons in the conduction band corresponding to the fraction of trivalent Tm atoms are responsible for conventional, predominantly AF, RKKY couplings which compete with the double-exchange interaction.

In the case of TmTe just above the semiconductor-to-metal transition, the valence state is still very close to 2+ ($v \approx 2.3$ at $P = 2.7$ GPa, from the lattice constant) and the above description in terms of Tm$^{2+}$ "impurities" in a trivalent matrix appears questionable. Instead, one might consider a symmetric model in which 4$f$-holes associated with trivalent sites would favor a ferromagnetic coupling among Tm$^{2+}$ moments. However, it should be kept in mind that this mechanism requires the possibility for the holes to propagate, which, in the case of $f$-electrons, cannot occur through direct $f$-band conduction but requires appreciable $f$-$d$ hybridization. Furthermore, in the limit of low carrier densities, the RKKY interaction is expected to have a positive sign [20], so that its effect should add to, rather than compete with, that produced by double-exchange. Indeed, such a contribution from RKKY interactions, initially FM then AF was invoked in early works on Eu$_{1-x}$Gd$_x X$ compounds [1] to explain, at least qualitatively, the variation of the paramagnetic Curie temperature with doping. In a subsequent work [4], it was objected that a more careful treatment of impurity states is required, especially for low Gd doping, to account for the anomalies observed in transport properties. In this sense, however, the case of TmTe appears to be simpler because the conducting state is achieved gradually without introducing impurity atoms. It thus seems reasonable to ascribe the FM behavior to a RKKY-type mechanism in the limit of a low carrier concentration. It can be remarked that the maximum of $T_C = 14$ K is much larger than the value expected from the de Gennes factor calculated for normal metallic $R^{3+}$Te compounds. The latter picture does not take into account the differences in carrier concentrations.

The observation of a linear correlation between the pressure variations of $T_C$ and $\mu_0^2$ is one of the most striking results of this work. It implies that, as the degree of valence mixing increases, the ordering temperature decreases primarily owing to a reduction of the magnetic moment. At 5.4 GPa, the value of the moment is already quite small (about the minimum value compatible with a Tm$^{2+}$ crystal-field eigenstate in cubic symmetry). In particular, it is much smaller than the AF moment ($\mu_{AF} = 1.7\mu_B$) found previously in stoichiometric TmSe. We believe that this quenching of the moment results from the enhancement of Kondo-type spin fluctuations. It is interesting to speculate that, for pressures of about 6 GPa, TmTe might develop no long-range order, thereby providing the first example of a non-magnetic mixed-valence state in thulium. The observation of a mean-field-type $T$-dependence of the order parameter, typical for the vicinity of a quantum critical point [21], lends some support to this idea. Neutron measurements below 1 K at higher pressures, as well as the use of a local probe such as Mössbauer effect, could help to solve this problem.

The authors wish to thank A. Gukasov for his help during the experiments and P. Haen for communication of unpublished results and useful discussion. This work has been supported by the Commission of the European Union through TMR grant Nr. ERB 400 1 GT 95 5456

---